\documentclass[aps,prx,twocolumn,superscriptaddress]{revtex4-2}
\usepackage{amsfonts}
\usepackage{graphicx}
\usepackage{epstopdf}
\usepackage{epsfig}
\usepackage{amsmath}
\usepackage[normalem]{ulem}
\usepackage{multirow}
\usepackage{makecell}
\usepackage{array}
\usepackage{booktabs}
\usepackage{mathtools}
\usepackage{bm}
\usepackage{color}
\usepackage{soul,xcolor}
\usepackage{array}
\usepackage{booktabs}
\usepackage{braket}
\usepackage{mathrsfs}
\usepackage{dsfont}

\usepackage{tabularx}
\newcolumntype{Y}{>{\centering\arraybackslash}X}

\usepackage{blindtext}
\usepackage{comment}
\definecolor{lightblue}{rgb}{0.3, 0.3, 0.90}
\definecolor{darkblue}{rgb}{0.0, 0.2, 0.5}
\definecolor{lightred}{rgb}{1.0, 0.6, 0.6}
\definecolor{darkred}{rgb}{0.8, 0.0, 0.0}

\begin{document}

\title{Floquet quantum many-body scars in the tilted Fermi-Hubbard chain}

\author{Jun-Yin Huang}
\affiliation{School of Electrical, Computer and Energy Engineering, Arizona State University, Tempe, Arizona 85287, USA}

\author{Li-Li Ye}
\affiliation{School of Electrical, Computer and Energy Engineering, Arizona State University, Tempe, Arizona 85287, USA}

\author{Ying-Cheng Lai} \email{Ying-Cheng.Lai@asu.edu}
\affiliation{School of Electrical, Computer and Energy Engineering, Arizona State University, Tempe, Arizona 85287, USA}
\affiliation{Department of Physics, Arizona State University, Tempe, Arizona 85287, USA}

\date{\today}

\begin{abstract}

The one-dimensional tilted, periodically driven Fermi-Hubbard chain is a paradigm in the study of quantum many-body physics, particularly for solid-state systems. We uncover the emergence of Floquet scarring states, a class of quantum many-body scarring (QMBS) states that defy random thermalization. The underlying physical mechanism is identified to be the Floquet resonances between these degenerate Fock bases that can be connected by {\it one hopping process}. It is the first-order hopping perturbation effect. Utilizing the degenerate Floquet perturbation theory, we derive the exact conditions under which the exotic QMBS states emerge. Phenomena such as quantum revivals and subharmonic responses are also studied. Those results open the possibility of modulating and engineering solid-state quantum many-body systems to achieve nonergodicity.

\end{abstract}
\maketitle


\section{Introduction} \label{sec:intro}

Since the experimental observation of quantum revivals in Rydberg atom 
arrays~\cite{bernien2017probing}, the phenomenon of quantum many-body scarring 
(QMBS)~\cite{serbyn2021quantum} has attracted a great deal of interest~\cite{turner2018weak,turner2018quantum,PhysRevB.100.184312,PhysRevLett.122.040603,PhysRevLett.122.173401,bull2020quantum,PhysRevResearch.2.033044,PhysRevX.11.021021,PhysRevB.98.235155,PhysRevB.98.235156,PhysRevB.101.195131,PhysRevB.102.085120,PhysRevB.102.085140,hudomal2020quantum,desaules2021proposal,PhysRevLett.130.250402}. In general, 
QMBS states signify a weak breaking of ergodicity and thus a violation of the 
eigenstate thermalization hypothesis (ETH)~\cite{Deutsch:1991,Srednicki:1994} for 
quantum many-body interacting systems that are expected to thermalize and thus be 
ergodic~\cite{RDO:2009}. A recent experimental work~\cite{bluvstein2021controlling} 
showed that quantum revivals can be enhanced and stabilized via periodic driving, 
opening the possibility that QMBS can arise in quantum Floquet systems and raising
the questions of whether QMBS states can arise in driven quantum systems in general.
An affirmative answer would open the door to exploiting Floquet engineering for 
modulating and controlling the QMBS dynamics, and uncovering the underlying physical 
mechanism responsible for the emergence of Floquet scarring states then becomes an 
important issue. There were recent efforts in systems such as the driven PXP 
model~\cite{mukherjee2020collapse,PhysRevResearch.2.033284,PhysRevResearch.3.L012010,PhysRevB.102.075123,hudomal2022driving,huang2024engineering}, 
the Bose-Hubbard model~\cite{PhysRevLett.124.160604,PhysRevB.102.014301,PhysRevResearch.5.023010,beringer2024controlling}, 
diecrete-time crystals~\cite{maskara2021discrete,PhysRevLett.129.133001,huang2023analytical},
and others~\cite{PhysRevLett.123.136401,haldar2021dynamical,rozon2022constructing,ljubotina2024tangent}.
For example, in the PXP models under some engineered driving protocols, a breakdown 
of the ETH was demonstrated and the Floquet scarring states were analyzed~\cite{mukherjee2020collapse,PhysRevResearch.2.033284,PhysRevResearch.3.L012010}. 
Most existing works on the Floquet scarring dynamics were based on the PXP model with 
engineered driving protocols. 

The one-dimensional (1D) Fermi-Hubbard chain represents another paradigm for studying 
complex many-body physics, particularly in solid-state systems. Recently, experimental 
realization of the 1D titled Fermi-Hubbard chain was achieved by using cold atoms in 
a 3D optical lattice~\cite{scherg2021observing}, providing a natural setting for 
investigating weak ergodicity breaking due to Hilbert space 
fragmentation~\cite{scherg2021observing,PhysRevLett.130.010201,papic2022weak}. It was
also found that, beyond fragmentation, the 1D titled Fermi-Hubbard chain hosts QMBS
states in some specific regime at half filling~\cite{desaules2021proposal}. An 
outstanding question is whether Floquet scarring states can generally arise in the 
driven tilted Fermi-Hubbard systems. We note that, if the answer is affirmative, the 
cold-atom systems would provide a feasible experimental platform for verification, 
where the on-site Coulomb interaction strength can be readily controlled through a Feshbach 
resonance~\cite{courteille1998observation,chin2010feshbach,zhao2020quantum}. Another
potential experimental system is the lattices of dopant-based quantum 
dots~\cite{WKFWNKRBS:2022}. The Floquet tilted Fermi-Hubbard chain is a suitable choice, not only for its fundamental role as a many-body model but for its potential in direct experimental verification, complementing previous studies focused on the PXP model.

In this paper, we aim to uncover Floquet scarring states in 1D tilted 
Fermi-Hubbard chain with periodically driven on-site Coulomb interaction. We first 
numerically identify the signatures of the possible Floquet scarring states 
according to the typical features of QMBS states in the static 
chain~\cite{desaules2021proposal}, which include persistent quantum revivals 
following quenches from some specific initial states, suppressed entanglement 
entropy, and the scarred tower structures in the overlaps of the Floquet eigenstates 
with some specific initial states. We find that the emergence of possible 
Floquet scarring dynamics is associated with robust synchrony of the quantum 
state with the driving frequency, regardless of its strength. In particular, 
the scarring dynamics periodically emerges as the static detuning term of the Coulomb
interaction varies in integer multiples of the driving frequency. Exploiting the 
degenerate Floquet perturbation theory~\cite{mukherjee2020collapse}, we derive the 
analytic emergence conditions for the Floquet scarring states. It leads to the underlying mechanism: the Floquet scarring dynamics are the 
results of the resonances between these degenerate Fock bases that can be connected by {\it one hopping process}. In other words, the resonances induced by first-order hopping perturbation lead to the Floquet scarring dynamics.

In general, the resonances may lead to unbounded heating in many-body Floquet systems, thus a stable scarring state requires the absence of resonances~\cite{haldar2021dynamical}. Here the resonance mechanism uncovered is surprising, making it possible to ``heat up'' the system in a nonergodic manner. In Ref.~\cite{haldar2021dynamical} on Ising and Heisenberg interacting systems, it was found that the resonances play a somewhat opposite role in the emergence of Floquet scarring states. 
In these systems, the emergence mechanism was found to be dynamical freezing under 
a strong driving. At the so-called ``scar points'', the longitudinal magnetization 
becomes an emerged conserved quantity, preventing the system from heating up 
ergodically - the phenomenon of freezing. The resonances tend to destroy the 
inertness of the ``scar point'', implying the emergence of stable Floquet scarring 
dynamics without resonances. A similar role of resonances also was observed in the driven PXP model~\cite{mukherjee2020collapse}. The reason for the seeming contradiction with our work lies in the nature of the unperturbed dynamics. In their system, the 
unperturbed systems can heat up ergodically, which is thermal. At ``scar points'', 
the dynamics are severely constrained by the emergence of the local conserved 
quantity, while the resonances would significantly weaken the dynamical constraint. 
In our study, the unperturbed system does not thermalize because all the fermions 
are fully confined to their initial lattice sites. The resonances induced by the 
hopping perturbation then open the way to heat up. In addition, the resonances do 
not lead to unbounded heating, since the hopping amplitude is typically much smaller 
than on-site Coulomb interaction and tilted potential strength.

\begin{figure} [t!]
\includegraphics[width=\linewidth]{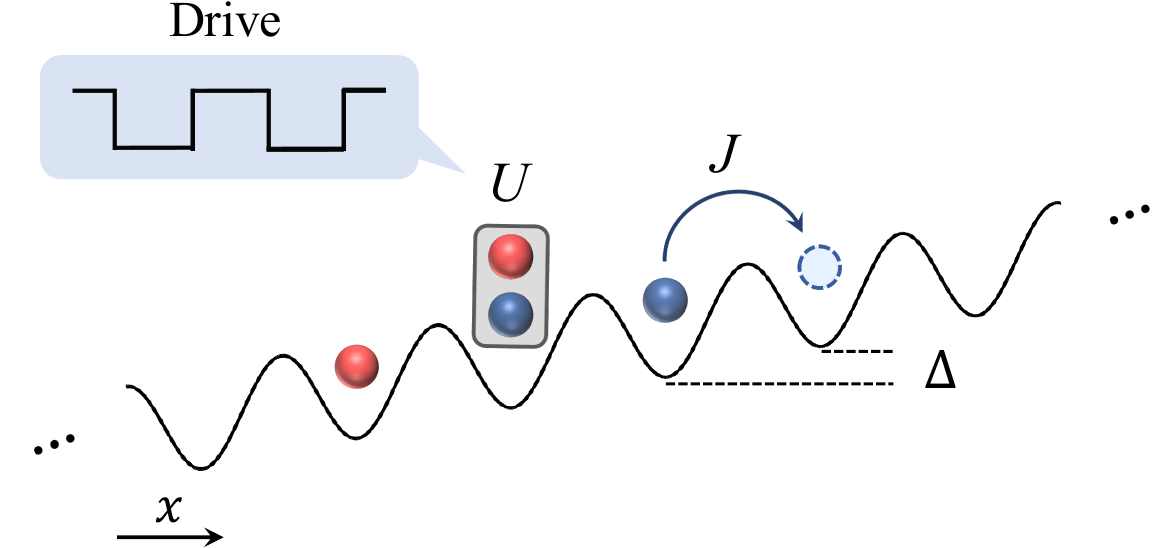}
\caption{A schematic illustration of the 1D tilted Fermi-Hubbard chain. The on-site
Coulomb interaction is driven by a periodic signal: 
$U(t)= U_0+U_m \ {\rm sgn} \left(\cos(\omega t)\right)$, where $J$ is the 
nearest-neighbor hopping amplitude and $\Delta$ is a spin-independent tilted 
potential. The spin up (down) fermions are colored in red (blue).}
\label{fig:1}
\end{figure}

We also find that, similar to the static chain, the equal quasienergy separation 
of the scarred towers is responsible for the observed quantum revivals~\cite{mukherjee2020collapse,PhysRevResearch.5.023010}. The subharmonic 
and incommensurate responses of the revivals to driving are observed in distinct 
frequency regimes. These responses and the synchronization effect open the door 
to modulating and engineering the Floquet scarring 
dynamics~\cite{hudomal2022driving,huang2024engineering}. 

In Sec.~\ref{sec:model}, we introduce the 1D driven tilted Fermi-Hubbard chain and describe the phenomenon of QMBS in the corresponding static chain. The Floquet scarring states 
are investigated in Sec.~\ref{sec:floquetscar}, where the phenomenon of quantum 
revivals is studied in Sec.~\ref{subsec:revivals} and the conditions dictating the 
emergence of the Floquet scarring states are obtained numerically in 
Sec.~\ref{subsec:emergence}. An analytic derivation of the emergence conditions
is presented in Sec.~\ref{sec:theory}, based on the degenerate Floquet perturbation
theory. And the connection between the driven and Floquet QMBS dynamics is discussed in Sec.~\ref{sec:connection}. The phenomena of subharmonic and incommensurate responses to driving are presented in Sec.~\ref{sec:responses}. A summary and discussion are presented in Sec.~\ref{sec:discussion}. The methods for calculating the quantum evolution dynamics, bipartite von Neumann entanglement entropy, and an error analysis are given in Appendix~\ref{appendix:A} and the transition from Wannier-Stark localization to Floquet scar phase is described in Appendix~\ref{appendix:B}. The detailed introductions for the Floquet perturbation theory are in Appendix~\ref{appendix:C}, and the robust period-doubling phenomenon is shown in Appendix~\ref{appendix:D}.

\section{1D Tilted Fermi-Hubbard chain} \label{sec:model}
The 1D tilted Fermi-Hubbard chain under periodic driving is given by Hamiltonian~\cite{scherg2021observing,desaules2021proposal}
\begin{align} \label{eq:H}
    H = & \sum_{j,\sigma=\uparrow,\downarrow} \left(-J\hat{c}_{j,\sigma}^{\dag}\hat{c}_{j+1,\sigma}+{\rm h.c.}+\Delta j\hat{n}_{j,\sigma}\right)\notag \\
    & +U(t)\sum_{j}\hat{n}_{j,\uparrow}\hat{n}_{j,\downarrow},
\end{align}
where $\hat{c}_{j,\sigma}^{\dag}$ ($\hat{c}_{j,\sigma}$) is the fermionic creation 
(annihilation) operator on site $j$ with the spin index $\sigma$, 
$\hat{n}_{j,\sigma}=\hat{c}_{j,\sigma}^{\dag}\hat{c}_{j,\sigma}$ is the density 
operator, $J$ and $\Delta$ are the nearest-neighbor hopping amplitude and 
spin-independent tilted potential, 
respectively. For simplicity, the on-site Coulomb interaction is defined with a square-wave driving function: $U(t)= U_0+U_m \ {\rm sgn} \left(\cos(\omega t)\right)$, where $U_0$ is the static detuning, $U_m$ is the modulation amplitude, and $\omega$ is the driving frequency. The linear static tilt $\Delta$ can be implemented using a magnetic 
field gradient and the time-periodic signal $U(t)$ can be modulated via a 
Feshbach resonance~\cite{courteille1998observation,chin2010feshbach,zhao2020quantum}. 
Based on the previous work~\cite{desaules2021proposal}, the system is proposed to consist of an even number $L$ of sites, with the initial state containing equal numbers of spin-up and spin-down fermions. Periodic boundary conditions are applied to eliminate boundary effects.

To recognize Floquet scarring states, we first describe QMBS states in the 
corresponding undriven system~\cite{desaules2021proposal}. We use the following notations: $\uparrow$ for spin 
up, $\downarrow$ for spin down, $0$ for an empty site, and $\updownarrow$ for a 
doublon. At the filling factor 
\begin{align}
\nu = (N_{\uparrow}+ N_{\downarrow})/L = 1, 
\end{align}
the undriven system hosts QMBS states in the regime $\Delta \approx U \gg J$, 
which can be conveniently probed using a quantum quench 
process from some special non-equilibrium initial states $\ket{\psi_s}$. Such initial 
states can be
\begin{align} \nonumber
	\ket{\downarrow \uparrow \uparrow \downarrow \cdots} \ \mbox{and} \ 
\ket{\downarrow \cdots \downarrow \updownarrow 0 \uparrow \cdots \uparrow}, 
\end{align}
as well as their spin-reversed states 
\begin{align} \nonumber
	\ket{\uparrow \downarrow \downarrow \uparrow \cdots} \ \mbox{and} \ 
\ket{\uparrow \cdots \uparrow \updownarrow 0 \downarrow \cdots \downarrow}.
\end{align}
A salient feature of QMBS states is the fidelity revival observed, during the time evolution that starts from the special initial states $\ket{\psi_s}$. The fidelity is represented by the overlap between the time-evolved quantum state $\ket{\psi(t)}$ and its initial state $\ket{\psi_0}$, which is defined as
\begin{align} \label{eq:fidelity}
    F(t) \equiv |\!\braket{\psi(t)|\psi_0}\!|^2.
\end{align}
However for quenches from other non-equilibrium initial states denoted as $|\psi_{th}\rangle$, the undriven system typically thermalizes quickly, in such cases, the fidelity $F(t)$ rapidly decays to zero and remains near zero over time. 

\begin{figure} [ht!]
\includegraphics[width=0.95\linewidth]{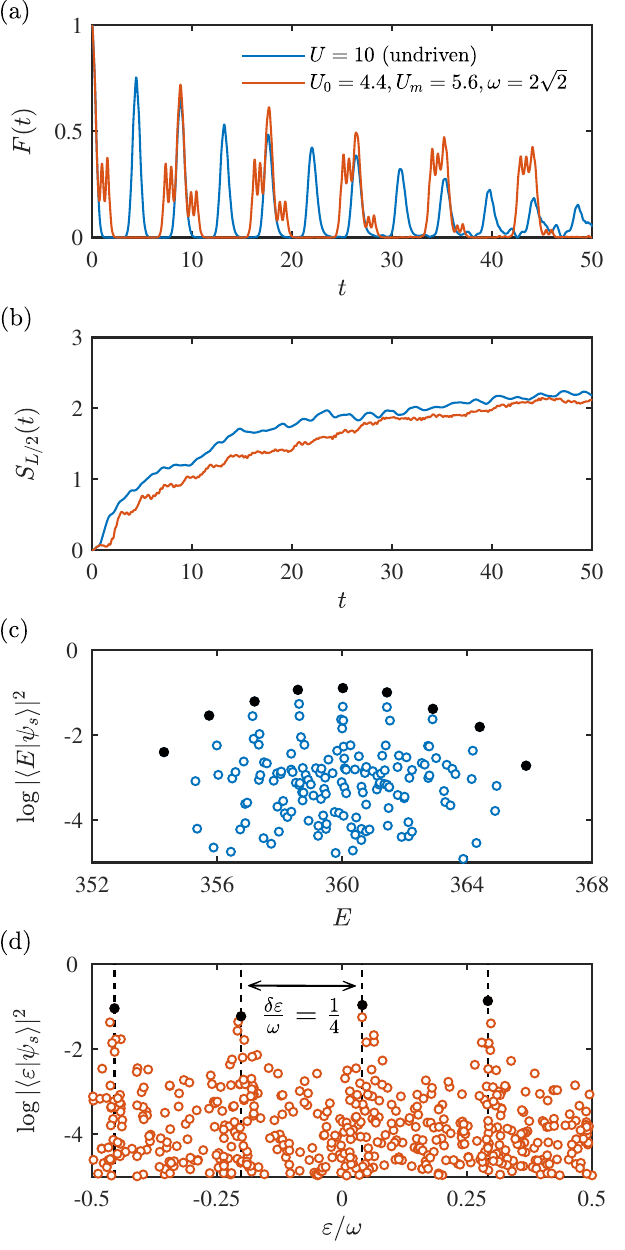}
\caption{Scarring dynamics in a quench process in the 1D tilted Fermi-Hubbard system. 
The initial state is 
$\ket{\psi_s} = \ket{\downarrow \uparrow \uparrow \downarrow \downarrow \uparrow \uparrow \downarrow}$. 
The system parameters are $L=8$ and $\Delta = 10$. The undriven case for $U=10$ is
represented by the blue color, and the driven case by orange for $U_0=4.4$, 
$U_m = 5.6$, and $\omega = 2\sqrt{2}$. Time evolution of: (a) wave function 
fidelity $F$ and (b) bipartite entanglement entropy $S_{L/2}$. (c-d) The overlap of 
the eigenstates and Floquet eigenstates with $\ket{\psi_s}$ for the undriven and driven cases, 
respectively, where the black dots indicate the top of every tower structures, 
corresponding to the scarring states in (c) and the Floquet scarring states in (d). 
These towers have an equal or approximately equal energy separation of about 
$\sqrt{2}$ in (c) and $\omega/4$ in (d).} 
\label{fig:2}
\end{figure} 

Differing from previous work~\cite{desaules2021proposal}, we treat the full chain 
directly, following the numerical methods in Ref.~\cite{scherg2021observing}. The 
details are provided in Appendix~\ref{appendix:A}. For convenience, the hopping parameter and the Planck constant are normalized to $J\equiv 1$ and $\hbar\equiv 1$, respectively. And the site number $L = 8$ is considered. In an undriven chain, the revivals from the initial state 
$\ket{\psi_s} = \ket{\downarrow \uparrow \uparrow \downarrow \downarrow \uparrow \uparrow \downarrow}$ are shown in blue color in Fig.~\ref{fig:2}(a), where the revival period 
is $T_* \approx \sqrt{2}\pi$. The revivals are not perfect, where the height of the revival peak decreases with time. Another quantity characterizing the evolution of a quantum state is the bipartite von Neumann entanglement entropy $S_{N/2}$, which is 
suppressed in a quantum quench. Figure~\ref{fig:2}(b) plots 
$S_{L/2} = S_{l} = -{\rm tr} \rho_l \log{\rho_l}$ (blue), where the subscript 
$l$ ($r$) denotes the left (right) half-chain, and 
$\rho_l(t)={\rm tr}_r{\ket{\psi(t)}\!\bra{\psi(t)}}$ is the reduced density matrix 
for the left subsystem by tracing out the right subsystem. The system eigenstates 
can be calculated by diagnalizing the Hamiltonian of the full chain in the standard 
Fock space. The overlap of eigenstates with the initial state $\ket{\psi_s}$ is 
shown in Fig.~\ref{fig:2}(c), demonstrating the scarred 
eigenstates~\cite{turner2018weak} as marked by the scarred tower structures and the black 
dots at the top of the towers. These towers have a near-equal energy separation 
$\delta E \approx \sqrt{2}$, as the ``embedding'' construction in a thermal 
eigenstate. The scarred eigenstates have an abnormally high overlap with the 
initial state $\ket{\psi_s}$, resulting in the revivals in Fig.~\ref{fig:2}(a) with 
the revival period $T_* \approx 2\pi/\delta E$, i.e., $\omega_* \approx \delta E$.

\section{Emergence of Floquet scarring states} \label{sec:floquetscar}

\subsection{Quantum revivals} \label{subsec:revivals}

Figure~\ref{fig:2}(a) presents an example of the phenomenon of quantum revivals, 
where the fidelity exhibits distinct peaks during the time evolution and the 
revival period is about twice of that for the undriven case: $T_r \approx 2T_*$. 
For static detuning $U=U_0= 4.4$, there is no revival of the initial state 
$\ket{\psi_s}$ due to the rapid thermalization (not displayed in the figure). The results in Fig.~\ref{fig:2}(a) 
suggest that the periodic driving induces and enhances quantum revivals, as 
characterized by the higher revival amplitude in Fig.~\ref{fig:2}(a). Similarly,
the entanglement entropy $S_{L/2}$ in the driven system is relatively lower,
as shown in Fig.~\ref{fig:2}(b). 

Insights into the driven revival dynamics from $\ket{\psi_s}$ can be gained by 
studying the Floquet eigenstates. In particular, the periodically modulated 
Hamiltonian $H(t)=H(t+T)$ is determined by the time evolution of the Floquet 
operator over one period $T$~\cite{bukov2015universal}: 
\begin{equation}
    \mathcal{U}(t_0+T,t_0) = \mathcal{T} \exp{\left[ -\mathrm i \int_{t_0}^{t_0+T} H(t) dt \right]},
\end{equation}
where $\mathcal{T}$ denotes the time ordering and the initial time $t_0$ is set 
to $0$. For square-wave driving, the Floquet operator becomes
\begin{equation}
    \mathcal{U} = e^{-\mathrm i H_+ T/4} e^{-\mathrm i H_- T/2} e^{-\mathrm i H_+ T/4},
\end{equation}
where 
\begin{align}
H_{\pm} = H_{\rm s} \pm U_m \sum_{j}\hat{n}_{j,\uparrow}\hat{n}_{j,\downarrow} 
\end{align}
with the static detuning Hamiltonian $H_{\rm s}$. The Floquet operator is unitary 
with complex eigenvalues $\left\{e^{-\mathrm i \varepsilon_n T}\right\}$ and Floquet 
eigenstates $\left\{\ket{n}\right\}$. The quantities $\left\{\varepsilon_n \right\}$ 
are multi-valued, whereas the quasienergies 
$\left\{\varepsilon_n \ {\rm mod} \ \omega \right\}$ can be uniquely determined by a 
shift. Further, the time-independent stroboscopic Floquet 
Hamiltonian~\cite{bukov2015universal} $H_{\rm F}$ can be defined according to 
$\mathcal{U}=e^{-\mathrm i H_{\rm F} T}$, following 
$H_{\rm F} \ket{n} = \varepsilon_n \ket{n}$. The quasienergies and the Floquet 
eigenstates can be calculated through exact diagonalization of the Floquet 
operator $\mathcal{U}$ in the standard Fock space. For $L$ sites and filling 
factor $\nu = 1$, the dimension of this space is
\begin{align} \nonumber
\left(
\begin{matrix}
L \\ L/2 
\end{matrix}
\right)
\times  \left(
\begin{matrix}
L \\ L/2 
\end{matrix}
\right).
\end{align}
For $L=8$, the dimension is $4900$.

Figure~\ref{fig:2}(d) shows the overlap of the Floquet eigenstates with the initial 
state $\ket{\psi_s}$ for the same values of the driving parameters as in Fig.~\ref{fig:2}(a). The quasienergies fall within the interval $(-\omega/2,\omega/2)$ of the driving 
frequency, exhibiting four apparent tower structures with near-equal quasienergy 
separation $\delta \varepsilon \approx \omega/4 \approx \sqrt{2}/2$. The tops of 
these towers correspond to the Floquet scarring eigenstates, marked by the black 
dots. The strong overlaps are akin to the ones in Fig.~\ref{fig:2}(c). The equal 
quasienergy separation of the towers is responsible for quantum revivals: the quasienergy separation equals the revival frequency $\omega_r \approx \delta \varepsilon$ (the similar property is also noted in Refs. \cite{mukherjee2020collapse, PhysRevResearch.5.023010}). Combining the relation $\delta E = 2\delta \varepsilon$, it gives the doubling period $T_r \approx 2T_*$.

\begin{figure*} [ht!]
\includegraphics[width=\linewidth]{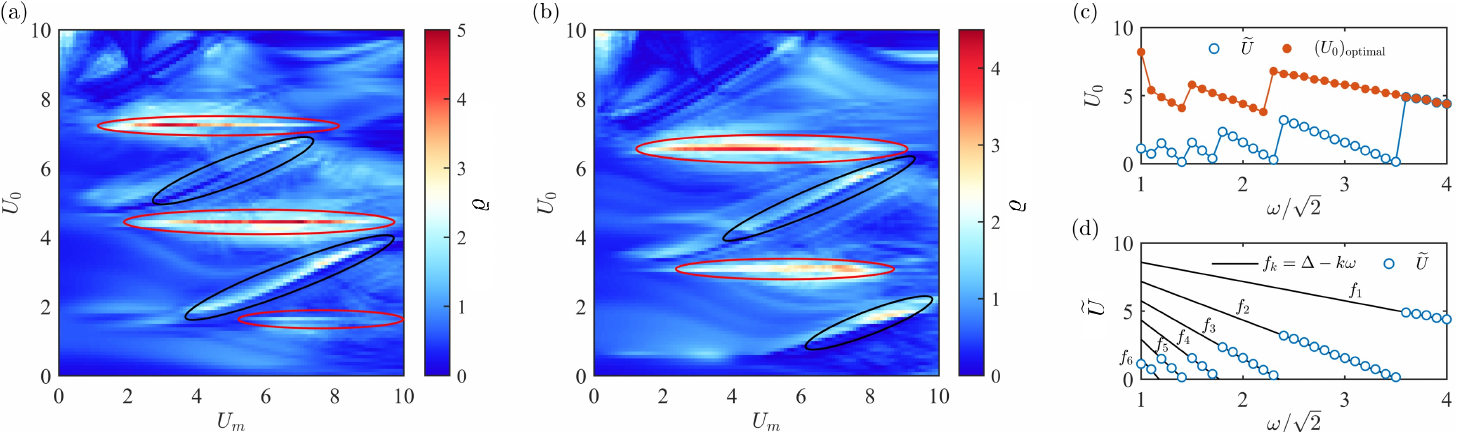}
\caption{Emergence of Floquet scarring in driven 1D tilted Fermi-Hubbard system. 
The system size is $L=8$. (a,b) Relative discrepancy $\varrho$ of the average 
fidelity between two initial states $\ket{\psi_s}$ and 
$\ket{\psi_{th}} = \ket{\uparrow \downarrow \uparrow \downarrow \uparrow \downarrow \uparrow \downarrow}$ 
in the parameter plane $(U_0,U_m)$. The average fidelity is calculated over the 
time interval $[0,50]$. The driving frequency is $\omega = 2\sqrt{2}$ for (a) and 
$\omega = 3.5$ for (b). The Floquet scarring states (encircled in red) appear at some 
specific values of $U_0$. The regions surrounded by the black 
curves correspond to the transition states. (c) The quantity $\widetilde{U}$ as the minimum threshold value for the emergence of the scarring dynamics (blue dots) and $(U_0)_{\rm optimal}$ corresponding to the 
maximal value of $\varrho$ (red dots) for 31 discrete values of the driving frequency.
(d) The quantity $\widetilde{U}$ characterized by a series of linear functions 
$f_k = \Delta -k \omega$, for $\Delta = 10$ and $k=1,2,\cdots, 6$.}
\label{fig:fid_ave}
\end{figure*}

\subsection{Emergence conditions of Floquet scarring states} \label{subsec:emergence}

To uncover the dependence of $\omega_r$ on the driving parameters, we first search 
for potential Floquet scarring states. In particular, we fix $\Delta = 10$ and scan 
the parameter plane of $U_0$ and $U_m$ to calculate the average fidelity for different 
driving frequencies: 
\begin{align} 
\langle F \rangle_t = \frac{1}{\tau} \int_0^{\tau} F(t) dt,
\end{align}
where the upper integration bound $\tau$ is set as $50$. In an approximate sense, the 
average fidelity characterizes the revivals. Note that a high value of the average 
fidelity is not necessarily indicative of revivals, as it may be the result of 
many-body localization or extremely slow thermalization. The following relative 
discrepancy of the average fidelity between different initial states provides a more 
appropriate way to characterize quantum revivals:
\begin{equation}
\varrho = \frac{\langle F_s \rangle_t - \langle F_{th} \rangle_t}{\langle F_{th} \rangle_t},
\end{equation}
where the subscripts $s$ and $th$ correspond to the initial state $\ket{\psi_s}$ and 
another one chosen as $\ket{\psi_{th}} = \ket{\uparrow \downarrow \uparrow \downarrow \cdots }$, 
respectively. The relative discrepancy $\varrho$ in fact quantifies the degree of quantum revivals after removing the thermal decay behavior of quench from $\ket{\psi_{th}}$.
Figures~\ref{fig:fid_ave}(a) and \ref{fig:fid_ave}(b) show $\varrho$ versus $U_0$ 
and $U_m$ for two values of the driving frequency: $\omega = 2\sqrt{2}$ and 
$\omega = 3.5$, respectively, for $L=8$. The regions with high $\varrho$ values are 
encircled in red, in which Floquet scarring states arise. The Floquet scarring states appear for some specific $U_0$ values (denoted as $U_0^s$) over a wide range of 
$U_m$, as indicated by the horizontal lines with bright red. The results suggest:
\begin{equation} \label{eq:U_0s}
    U_0^s \approx \widetilde{U}+n\omega,
\end{equation}
where $\widetilde{U}$ is the minimum threshold value for the emergence of the 
scarring dynamics for $n$ being an integer and $\widetilde{U}$ depends only on 
the driving type and its frequency $\omega$. The regions encircled by the black 
curves do not correspond to the Floquet scarring states, even though their $\varrho$ 
values are not too small. In fact, in these regions, states are in a transition 
from Wannier-Stark localization to the Floquet scarring phase, where both the 
quantum fidelity quenching from $\ket{\psi_s}$ and $\ket{\psi_{th}}$ have large 
average values and oscillations, whereas the evolution of $\ket{\psi_s}$ 
revives without reaching zero. More details of the transition states are presented in 
Appendix~\ref{appendix:B}. 

We further scan the independent parameter space of $U_0 \in [0,10]$ and $U_m \in [0,10]$ for 31 discrete driving frequencies 
$\omega = \sqrt{2},1.1\sqrt{2},1.2\sqrt{2},\cdots,4\sqrt{2}$. At each frequency, 
the optimal parameter 
\begin{equation}
    (U_0)_{\rm optimal} = \arg \max_{U_0,U_m} \{\varrho(U_0,U_m)\}
\end{equation} 
corresponds to the most distinct scarring dynamics in the entire parameter plane $(U_0, U_m)$, as shown in Fig.~\ref{fig:fid_ave}(c). $\varrho(U_0,U_m)$ means that the quantity $\varrho$ is a function of parameters $U_0$ and $U_m$, and the values of $\widetilde{U}$ are plotted based on Eq.~(\ref{eq:U_0s}). The threshold value $\widetilde{U}$ decreases to zero linearly with increased $\omega$ and then attains a larger value. The optimal parameter $(U_0)_{\rm optimal}$ has a similar behavior. The dependency of $\widetilde{U}$ on $\omega$ can be characterized 
by a series of linear functions: 
\begin{equation}
    \widetilde{U} = \Delta - k \omega,
    \label{eq:U_0}
\end{equation}
for $k = 1,2,3,\cdots$, as shown in Fig.~\ref{fig:fid_ave}(d). Since 
$\widetilde{U}$ is the minimum $U_0^s$ within the range $0 \le U_0 \le 10$, the 
integer $k$ can be determined by $0 \le \Delta - k \omega < \omega$ for specific 
driving frequency $\omega$.

The relations (\ref{eq:U_0s}) and (\ref{eq:U_0}) are the conditions for the 
emergence of the Floquet scarring states that emerge periodically over a wide 
range of the modulation amplitude $U_m$ as the static detuning term $U_0$ varies. 
This signifies a resonance induced by the periodic driving, whose frequency is 
exact the driving frequency. 

\section{Analytic derivation of the emergence conditions} 
\subsection{Emergence conditions}\label{sec:theory}
The emergence of the Floquet scarring states, as stipulated by the conditions
in Eqs.~(\ref{eq:U_0s}) and (\ref{eq:U_0}) are our main results. We now 
analytically derive these conditions  from the degenerate Floquet perturbation 
theory~\cite{soori2010nonadiabatic,mukherjee2020collapse}. To 
begin, we express the Hamiltonian (\ref{eq:H}) as $H(t) = H_0(t) + V$, where
\begin{align}
    H_0(t) & = \Delta \sum_{j,\sigma=\uparrow,\downarrow}  j\hat{n}_{j,\sigma} + U(t)\sum_{j}\hat{n}_{j,\uparrow}\hat{n}_{j,\downarrow}, \notag \\
    V & = -J\sum_{j,\sigma=\uparrow,\downarrow}\left(\hat{c}_{j,\sigma}^{\dag}\hat{c}_{j+1,\sigma}+{\rm h.c.}\right).
\end{align}
In the standard Fock basis, $H_0(t)$ is a diagonal matrix and commutes with itself at 
different times, $V$ is completely off-diagonal and can be regarded as a small 
time-independent perturbation due to the conditions $\Delta \gg J$ and 
$(U_0+U_m) \gg J$. 

For the unperturbated case, $H(t) = H_0(t)$. The Floquet eigenstates are simply the Fock bases $\ket{\rm F}$, following $H(t)\ket{{\rm F}_i} = E_i(t)\ket{{\rm F}_i}$ with index $i$ marking the $i$-th Fock basis. And the Floquet modes are~\cite{bukov2015universal}
\begin{equation}\label{eq:modes}
    \ket{{\rm F}_i(t)} = e^{-\mathrm i \int_0^t dt' E_i(t')} \ket{{\rm F}_i}.
\end{equation}
Note that for $t=0$, the Floquet modes are the Floquet eigenstates: $\ket{{\rm F}_i(0)} = \ket{{\rm F}_i}$. Intuitively, without the hopping perturbation $V$, the number of spin up (down) fermions at each site does not change with time, and the energy varies in synchrony with the drive. Thus, the dynamics are fully constrained.

For small $V$, the Floquet modes start to hybridize and deviate slightly from the unperturbated Floquet modes. Combining Eq. (\ref{eq:modes}), the Floquet mode $\ket{{\rm F}'_i(t)}$ can be expanded in the unperturbed eigenstates set
$\{\ket{{\rm F}_i}\}$~\cite{mukherjee2020collapse}:
\begin{align} \label{eq:main_slight}
    \ket{{\rm F}'_i(t)} =  e^{-\mathrm i \int_0^t dt' E_i(t')} \ket{{\rm F}_i} + \sum_{j\neq i} c_j(t) e^{-\mathrm i \int_0^t dt' E_j(t')} \ket{{\rm F}_j},
\end{align}
where $c_j(t) \ll 1$ is of order $J/\Delta$, for all $j\neq i$ and all $t$. The coefficients $c_j(t)$ exactly characterize the small deviations from the unperturbated Floquet modes. For the perturbated eigenstate $\ket{{\rm F}'_i}$ at $t=0$, we have~\cite{mukherjee2020collapse}
\begin{equation} \label{eq:main_undegenrated}
    c_j(0) = -\mathrm i \braket{{\rm F}_j|V|{\rm F}_i} \frac{\int_0^T dt e^{\mathrm i\int_0^t dt' [E_j(t')-E_i(t')]}}{e^{\mathrm i \int_0^T dt [E_j(t)-E_i(t)]}-1}.
\end{equation}
More details about Eq. (\ref{eq:main_undegenrated}) are in Appendix~\ref{appendix:C}. The analysis so far holds for nondegenerate states. It breaks down when degeneracy 
occurs under the condition:
\begin{equation} \label{eq:main_degenerated}
    e^{\mathrm i \int_0^T dt [E_j(t)-E_i(t)]} = 1.
\end{equation}

Suppose that there are $p$ unperturbated eigenstates degenerate with a certain Fock basis $\ket{{\rm F}_i}$, satisfying the 
condition (\ref{eq:main_degenerated}) for $\ket{{\rm F}_i}$. These $p$ Fock bases can be denoted as $\ket{{\rm F}_{ij}}$ with $j=1,2,\cdots,p$, and $\ket{{\rm F}_i} \equiv \ket{{\rm F}_{i0}} $, following $H_0(t)\ket{{\rm F}_{ij}}=E_{ij}(t)\ket{{\rm F}_{ij}}$ and $E_{i0}(t) = E_i(t)$. All of them form a degenerate set $\mathcal{D}_i = \left\{\ket{{\rm F}_{ij}}| j = 0, 1, \cdots, p\right\}$. In terms of the degenerate perturbation theory~\cite{sakurai1967advanced}, we can disregard the expansion on the other unperturbated eigenstates, then any states in the perturbated degenerate set $\mathcal{D}'_i$ is now given by
\begin{equation} \label{eq:main_psi}
    \ket{{\rm F}'_{ij}(t)} = \sum_{j = 0}^p c_j(t) e^{-\mathrm i \int_0^t dt' E_{ij}(t')} \ket{{\rm F}_{ij}}
\end{equation}
at $t=0$, where all $c_j(0)$ are of order $1$ (instead of order $J/\Delta$). As a result of first-order perturbation, the Floquet Hamiltonian $H_{\rm F}$ can be given by~\cite{mukherjee2020collapse}
\begin{equation}
    (H_{\rm F})_{jj'} = \frac{\braket{{\rm F}_{ij}|V|{\rm F}_{ij'}}}{T} \int_0^T dt e^{\mathrm i \int_0^t dt' [E_{ij}(t')-E_{ij'}(t')]},
    \label{eq:H_F}
\end{equation}
where $j, j'=0,1,\cdots, p$, and details are in Appendix~\ref{appendix:C}.

In general, the scarring states, as some ``embedding'' constructions in the thermal 
eigenstates, are the result of an anomalously high overlap with the initial state, 
shown as the top of the tower structures in Figs.~\ref{fig:2}(c) and \ref{fig:2}(d). For $L=8$, the special initial state $\ket{\psi_s} = \ket{\downarrow\uparrow\uparrow\downarrow\downarrow\uparrow\uparrow\downarrow}$ is one of the Fock bases, and we let this certain Fock basis $\ket{{\rm F}_i} = \ket{\psi_s}$. In the nondegenerate case, the overlap of the perturbated Floquet eigenstates with the initial state is 
\begin{equation*}
    |\!\braket{{\rm F}'_j|{\rm F}_i}\!|^2 = \begin{cases}
    1, & j = i \\
    |c_i(0)|^2, & j \neq i    \end{cases}.
\end{equation*} 
According to Eq.~(\ref{eq:main_undegenrated}), the overlap has an anomalously high value if and only if $j = i$, which does not allow the formation of scarred tower structures.

Consequently, the Floquet scarring states can arise only in the degenerate case. Any states in $\mathcal{D}'_i$ may have anomalously high overlaps with $\ket{{\rm F}_{i0}}$, forming the scarred tower structures. It requires
\begin{equation}
    \int_0^T dt [E_{ij}(t)-E_{ij'}(t)] = 2k\pi,
\end{equation}
where $k$ is an integer. 

Next we will discuss the degenerate set $\mathcal{D}_i$. Since $\ket{{\rm F}_{i0}}$ lacks doublon, its eigenenergy is $E_{i0} = \sum_{k=1}^L k\Delta$. Other Fock bases with the same eigenenergy ($E_{i0}$) must be degenerate with $\ket{{\rm F}_{i0}}$, whose number is
\begin{equation*}
   \left(
\begin{matrix}
L \\ L/2 
\end{matrix}
\right) - 1.
\end{equation*}

If $\mathcal{D}_i$ is entirely composed of the above $\bigl(\begin{smallmatrix}
L \\ L/2
\end{smallmatrix} \bigr)$ Fock states, then $H_{\rm F}$ is just the zero-matrix according to Eq. (\ref{eq:H_F}), since all $\braket{{\rm F}_{ij}|V|{\rm F}_{ij'}}$ terms are zero. Then $\ket{{\rm F}'_{ij}}$, as the eigenvalues of $H_{\rm F}$, cannot have anomalously high overlaps with $\ket{{\rm F}_{i0}}$. To ensure that $H_{\rm F}$ has non-zero elements, $\mathcal{D}_i$ must be extended. In this regard, the introduced Fock bases can be connected to $\ket{{\rm F}_{i0}}$ by {\it one hopping process}. If hopping $\ket{\uparrow\downarrow} \leftrightarrow \ket{\updownarrow 0} \leftrightarrow \ket{\downarrow \uparrow}$ are allowed, i.e., the Fock bases with one doublon are introduced and their common eigenenergy is $E_{ij}(t) = U(t)-\Delta+\sum_{k=1}^L k\Delta $. The degenerate condition now is
\begin{align}
\int_0^T dt [E_{i0}(t)-E_{ij}(t)] & = \int_0^T dt [\Delta -U(t)] \notag \\
& = (\Delta-U_0)T \notag \\
&= 2k\pi,
\end{align} 
in other words, $U_0 = \Delta -k\omega$, which is exactly the emergence conditions obtained 
from numerical calculations: Eqs.~(\ref{eq:U_0s}) and (\ref{eq:U_0}).

The above analysis provides a clear physical insight. In the presence of a small hopping process, the Floquet 
eigenstates start to hybridize and deviate slightly from the Fock bases. The small
deviations are characterized by $c_q(t)$ in Eq.~(\ref{eq:main_slight}), corresponding to 
the nondegenerate case. During the hybridization, the hopping between a series of 
degenerated unperturbed Floquet eigenstates can subject the system to heating up and 
exhibiting stable non-thermal Floquet eigenstates with an anomalously high overlap 
with the initial state. The conclusion is that the Floquet scarring dynamics originate from the resonances between these degenerate Fock bases that can be connected by {\it one hopping process}.

\begin{figure}[ht!]
\centering
\includegraphics[width = 0.6\linewidth]{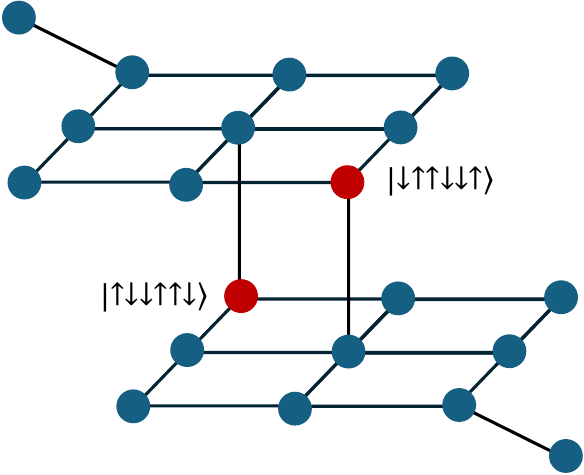}
\caption{Adjacency graph of effective Hamiltonian for undriven system with $L=6$ in the high tilt regime $\Delta \gg |U|, J$. The red vetices are $\ket{\psi_s}$: $\ket{\downarrow \uparrow \uparrow \downarrow \downarrow \uparrow}$ and $\ket{\uparrow \downarrow \downarrow \uparrow \uparrow  \downarrow}$. The blue vetices are other Fock states without doublon.
}
\label{fig:4}
\end{figure}

\begin{figure*}[ht!]
\centering
\includegraphics[width = 0.98\linewidth]{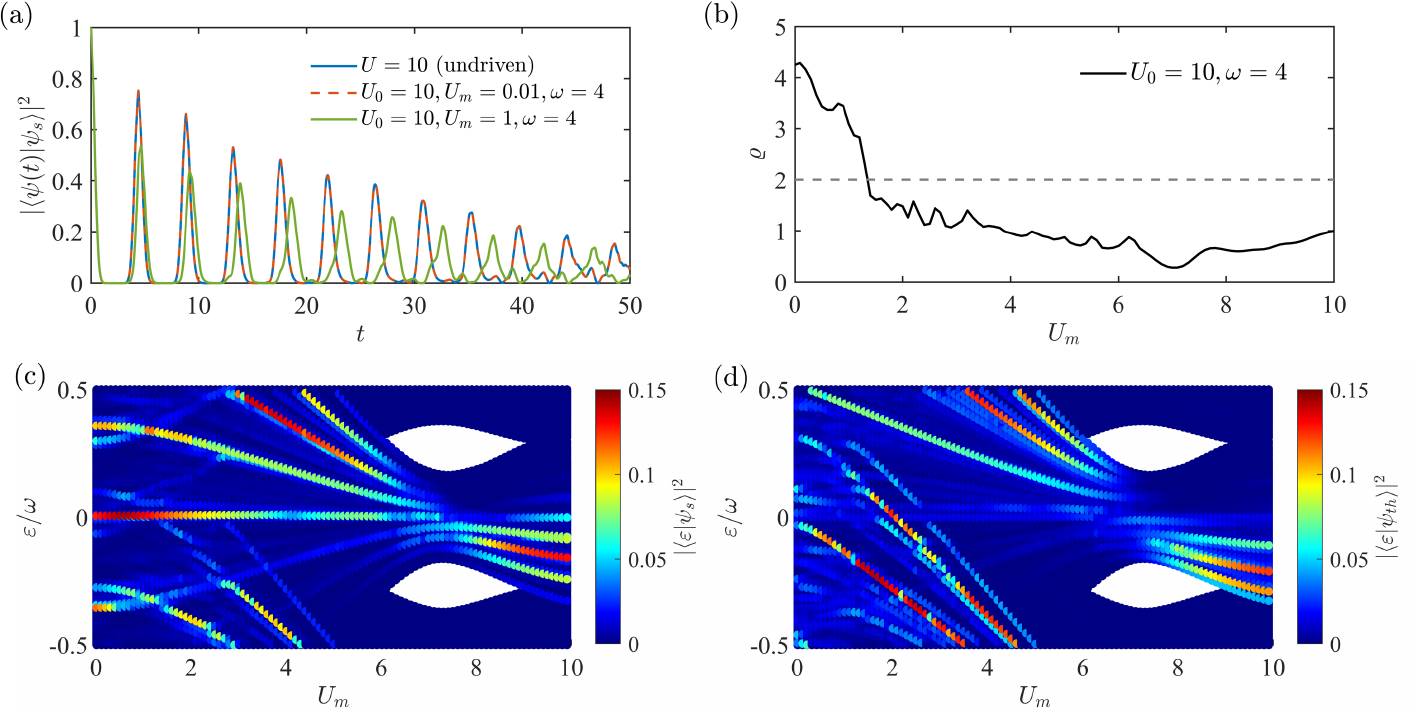}
\caption{From undriven scarring states to Floquet scarring states with driving parameters $U_0=10$ and $\omega=4$. (a) Time evolutions of wavefunction fidelity quenched from $\ket{\psi_s} = \ket{\downarrow \uparrow \uparrow \downarrow \downarrow \uparrow \uparrow \downarrow}$, for $U=10$ (blue), $U(t) = 10 + 0.01\ {\rm sgn} \left(\cos (4t)\right)$ (red), and $U(t) = 10 + 1 \ {\rm sgn}\left(\cos(4t)\right)$ (green). (b) Relative discrepancy $\varrho$ of the average fidelity between $\ket{\psi_s}$ and $\ket{\psi_{th}}$ as a function of  $U_m$. (c,d) Overlaps of the Floquet eigenstates with the initial state as $U_m$ varies from $0.01$ to $10$.
}
\label{fig:5}
\end{figure*}

\subsection{From undriven scarring states to Floquet one} \label{sec:connection}
Here, we would like to explore the connection between undriven and Floquet scarring states. For undriven system, scarring dynamics originates from a subgraph that is weakly connected to the rest of the Hamiltonian’s adjacency graph~\cite{desaules2021proposal}. The vertices of adjacency graph consist of a series of Fock states that share the same energy as the initial state $\ket{\psi_s}$. When the system is quenched from $\ket{\psi_s}$, the wave function $\ket{\psi(t)}$ slowly leaks out of this subgraph over time. In the regime $\Delta \approx U \gg J$, the effective Hamiltonian is
\begin{align}
    H_{\rm eff}^+ = & -J\sum_{j,\sigma=\uparrow,\downarrow} \hat{c}_{j,\sigma}^{\dag}\hat{c}_{j+1,\sigma}\hat{n}_{j,\overline{\sigma}}(1-\hat{n}_{j+1,\overline{\sigma}})+{\rm h.c.} \notag \\ 
    & +(U-\Delta)\sum_{j}\hat{n}_{j,\uparrow}\hat{n}_{j,\downarrow},
\end{align}
where hopping to the left is only allowed if it increases the number of doublons. This dynamical confinement causes the weakly connected subgraph. From the perspective of perturbation theory, the degenerate set $\mathcal{D}_i$ of $\ket{\psi_s}$ constitutes the vertices in adjacency graph, and each edge connecting two vertices represents allowed {\it one hopping process}.

Similar processes occur the corresponding Floquet system, however, the adjacency graph will alternate over time, due to the driving amplitude alternating between $U_0+U_m$ and $U_0-U_m$ over time. Taking driving protocol $U(t) = 4.4 + 5.6\ {\rm sgn} \left(\cos \left(2\sqrt{2} t\right) \right)$ and $L=6$ as an example, the adjacency graph remains the same as the undriven one when $t<T/4$ or $t>3T/4$, and when $T/4 \le t \le 3T/4$, it is in the high tilt regime $\Delta \gg |U_0-U_m|, J$ with effective Hamiltonian~\cite{scherg2021observing}
\begin{align}
    H_{\rm eff}^- = & J^{(3)}\hat{T}_3 + 2J^{(3)}\hat{T}_{XY} + 2J^{(3)}\sum_{j,\sigma}\hat{n}_{j,\sigma}\hat{n}_{j+1,\overline{\sigma}} \notag \\
    & + (U_0-U_m)\left(1-\frac{4J^2}{\Delta^2}\right)\sum_{j}\hat{n}_{j,\uparrow}\hat{n}_{j,\downarrow},
    \label{eq:effective}
\end{align}
where $J^{(3)} = (U_0-U_m)J^2/\Delta^2$ and 
\begin{align*}
    \hat{T}_3 & = \sum_{j,\sigma} \hat{c}_{j,\sigma}\hat{c}_{j+1,\sigma}^{\dag}\hat{c}_{j+1,\overline{\sigma}}^{\dag}\hat{c}_{j+2,\overline{\sigma}} + {\rm h.c.}, \\
    \hat{T}_{XY} & = \sum_{j,\sigma} \hat{c}_{j,\overline{\sigma}}^{\dag}\hat{c}_{j+1,\overline{\sigma}}\hat{c}_{j+1,\sigma}^{\dag} \hat{c}_{j,\sigma}.
\end{align*}
In this case, all Fock states without doublon constitute the vertices of the adjacency graph, as shown in Fig.~\ref{fig:4}. According to the Floquet theory, the effective adjacency graph is described by Floquet Hamiltonian $H_{\rm F}$, following $e^{-\mathrm i H_{\rm F} T} = e^{-\mathrm i H_{\rm eff}^+ T/4} e^{-\mathrm i H_{\rm eff}^- T/2} e^{-\mathrm i H_{\rm eff}^+ T/4}$. Due to $[H_{\rm eff}^+, H_{\rm eff}^-]\neq 0$, it needs to be solved by Eq. (\ref{eq:H_F}) within the framework of degenerate Floquet perturbation theory. The degenerate set $\mathcal{D}'_i$ consists of $\left(\begin{smallmatrix}L \\ L/2\end{smallmatrix}\right)=20$ Fock bases without doublon, $30$ Fock bases with one doublon $\ket{\updownarrow 0}$ segment, $12$ Fock bases with two doublon $\ket{\updownarrow 0}$ segments, and $\ket{\updownarrow 0\updownarrow 0\updownarrow 0}$. Thus the effective adjacency graph is similar as the undriven one, and the driving parameters $(U_0, U_m, \omega)$ determine the weights of the edges according to Eq. (\ref{eq:H_F}).

For the Floquet QMBS states, the emergence conditions are given by $U_0 = \Delta - k\omega$, where $k$ is an integer. In the limit of $U_m \to 0$, the Floquet dynamics converges to the undriven dynamics with $U = U_0$, regardless of the value of $\omega$, as as illustrated by the blue and red curves in Fig.~\ref{fig:5}(a). In this regard, the undriven QMBS states can be viewed as a special emergence at $U_0=\Delta=10$ and $k=0$. Then we provide some numerical illustrations from undriven scarring states to Floquet one. For $U_0 = 10$ and $U_m$ is in a range close to zero, the Floquet scarring states still persist, in agreement with our emergence condition. The variation of $\varrho$ as a function of $U_m$ is presented in Fig.~\ref{fig:5}(b), where a high $\varrho$ (such as the criterion $\varrho > 2$) signifies a pronounced revival. Roughly, the Floquet scarring states arise at $U_m \in [0, 1.3]$. When $U_0=10$ and $\omega = 4$, Figs.~\ref{fig:5}(c) and \ref{fig:5}(d) plot the deformations of the overlaps $|\langle \varepsilon|\psi_s \rangle|^2$ and  $|\langle \varepsilon|\psi_{th} \rangle|^2$, respectively, as $U_m$ increases from $0.01$ (undriven) to 10. The three highest overlaps $|\langle \varepsilon|\psi_s \rangle|^2$ in $U_m = 0.01$ continuously decrease as $U_m$ increases, and still remain at a relatively high level in the range of $U_m \in [0, 1.3]$ manifesting the persistence of Floquet scarring states.

The emergence condition induces the resonances between vertices in adjacency graph facilitating weak ergodicity breaking, and parameters $(J,\Delta,\nu,U_m,\omega)$ determine the edges and weights between different vertices thereby influencing the Floquet QMBS dynamics. Thus, the Floquet QMBS states include but extends far beyond the undriven one, and our emergence conditions offer a novel and profound perspective on both Floquet and undriven QMBS dynamics.

\section{Subharmonic and incommensurate responses} \label{sec:responses}

Figures~\ref{fig:2}(a) and \ref{fig:2}(d) show a fourth subharmonic response, a 
phenomenon first reported in the discrete time crystal~\cite{maskara2021discrete}, 
where the driven revival frequency is a quarter of the driving frequency: 
$\omega_r \approx \omega/4$. In the 1D PXP 
model~\cite{bluvstein2021controlling,hudomal2022driving}, under a driven chemical 
potential, when the initial state is the N${\rm \acute{e}}$el state, a robust 
(second) subharmonic locking of the scarring frequency $\omega_r \approx  \omega/2$ 
arises over a wide range of the driving frequency~\cite{bluvstein2021controlling}. 
In fact, the driven revival frequency is a function of $\omega$, $U_0$, and $U_m$, 
including harmonic, subharmonic, the forth subharmonic, etc., and even incommensurate 
responses. From the point of view of control and modulation, this implies a high 
degree of tunability.

\begin{figure} [ht!]
\includegraphics[width=\linewidth]{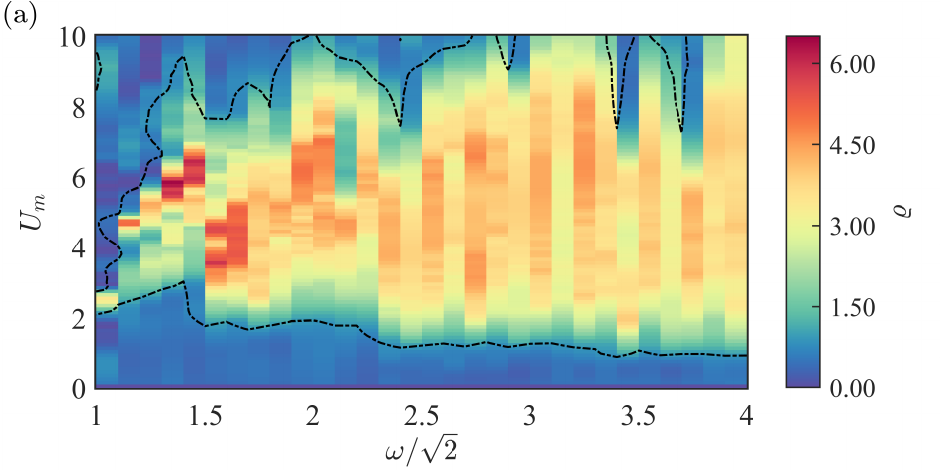}
\includegraphics[width=\linewidth]{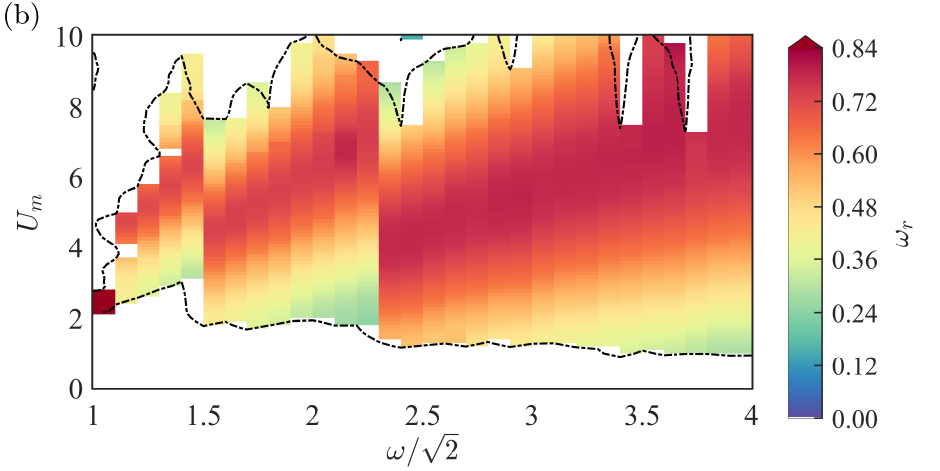}
\includegraphics[width=\linewidth]{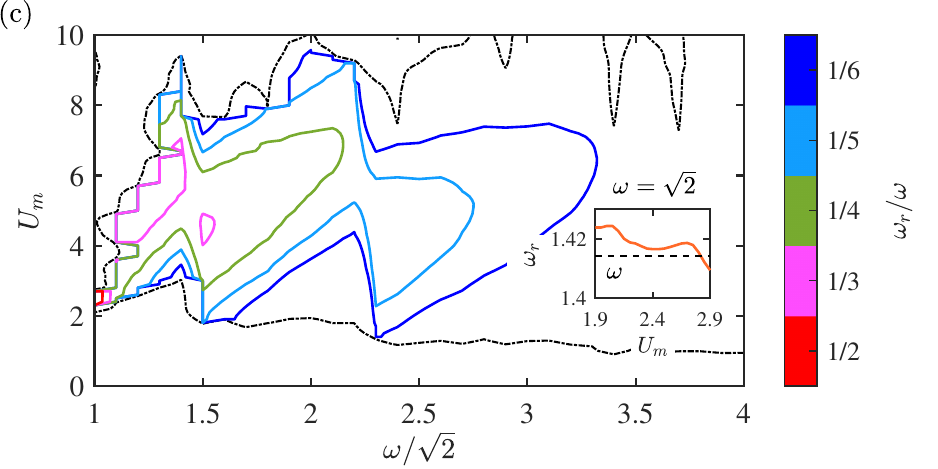}
\caption{Quantum revival properties of scarring dynamics in the driven 1D tilted 
Fermi-Hubbard systems. The emergence of the scarring states depends on the modulation 
amplitude $U_m$ and the driving frequency $\omega$. The system parameters are $L=8$ 
and $U_0 = (U_0)_{\rm optimal}$. The color scales indicate (a) the relative 
discrepancy $\varrho$, (b) the revival frequency $\omega_r$, and (c) the orders of 
subharmonic response.}
\label{fig:scan}
\end{figure}

We examine the parameter plane $(\omega,U_m)$ for the driven revival frequency at 
$U_0=(U_0)_{\rm optimal}$, which can be obtained as 
$\omega_r = \arg\max_{\omega}{[f(\omega)]}$, where
\begin{equation}
    f(\omega)=  \int_0^{\tau} F(t) e^{-\mathrm i \omega t} dt
\end{equation}
is the Fourier transform of $F(t)$ (we set $\tau = 100$ in numerical calculation). 
Figure~\ref{fig:scan}(a) shows the relative discrepancy $\varrho$ as a function of 
$\omega$ and $U_m$ for $U_0=(U_0)_{\rm optimal}$. In the frequency domain, a higher 
amplitude $f(\omega_r)$ always corresponds to narrower broadening at $\omega_r$, 
indicating higher revival peaks and more stable revival frequency, suggesting that
the value of $f(\omega_r)$ can be used to characterize the strength of the quantum 
revivals. The contour line of $f(\omega_r)=1$ is plotted 
in black chain curve. The frequency of the undriven revivals, $f(w_*) = 16.12$, serves as a reference point.

Figure~\ref{fig:scan}(b) shows the actual dependence of the revival frequency 
$\omega_r$ on $\omega$ and $U_m$ for $U_0=(U_0)_{\rm optimal}$, where the regions 
with high $\varrho$ correspond to the typical scarring dynamics. The regions with 
low $\varrho$ values ($\varrho < 1$) can then be disregarded, shown as the blank 
area with boundaries marked by the black chain curves. As $(U_0)_{\rm optimal}$ 
abruptly changes its value at $\omega/\sqrt{2}= 1.1$, $1.5$, and $2.3$ 
[Fig.~\ref{fig:fid_ave}(c)], the changes in $\omega_r$ are discontinuous at these 
driving frequencies. The modulation amplitude $U_m$ tends to shift towards a larger 
value when $(U_0)_{\rm optimal}$ switches to a larger value. For $\omega = \sqrt{2}$, 
the scarring region follows $\omega_r \ge 0.84$ as marked by the same color (deep 
red). Figure~\ref{fig:scan}(c) shows the contour lines representing a commensurate 
relation between $\omega_r$ and $\omega$, including the second, third, $\ldots$, and 
sixth subharmonic responses, i.e., $\omega_r= \omega/k$ with $k = 2, 3, \cdots, 6$, 
marked in different colors with the harmonic response ($\omega_r = \omega$) at 
$\omega = \sqrt{2}$ shown in the subgraph. An incommensurate relation can be 
realized in the regions between the adjacent contour lines. A convenient method to 
regulate these responses is fixing the driving frequency $\omega$ (with the 
corresponding $U_0$) and then tuning $U_m$, the so-called engineering subharmonic 
response via Floquet scarring states~\cite{huang2024engineering}. 

\section{Discussion} \label{sec:discussion}

In complex quantum systems, many-body interactions naturally lead to thermalization 
that destroys the coherence of the quantum states. However, QMBS states represent an
exception with significant potential applications, e.g., in quantum information 
science and technology. The phenomenon of QMBS has attracted a great deal of recent 
attention. From an application perspective, driven systems are of particular interest 
because of the possibility of realizing quantum control and engineering through some 
external driving input. In a periodically driven system, the QMBS states become the 
Floquet scarring states that have mostly been investigated using the PXP model that 
is specific to the Rydberg atomic systems. A field in which many-body interactions 
are fundamental is solid-state systems that are often more accessible to control and 
device engineering, rendering useful and important studying the phenomenon of Floquet 
scarring in these systems. A paradigm for probing into Fermionic many-body physics 
in these systems is the 1D Fermi-Hubbard chain.

We studied the 1D tilted Fermi-Hubbard system under a periodic driving. The 
corresponding static chain hosts QMBS states in a typical parameter regime. The 
scarring dynamics follow a quench from some special initial states and their 
spin-reversed states. Our computations and analysis provided unequivocal evidence 
for the emergence of the Floquet scarring states in the systems with physical 
manifestations including persistent quantum revivals, suppressed entanglement 
entropy, and the scarred tower structures in the overlaps of Floquet eigenstates 
with the initial state. A unique feature of the towers is that they have an equal 
quasienergy separation that is approximately the revival frequency. This feature is associated with the wave function fidelity undergoing a constructive (or destructive) process to reach the local maximum (or minimum), similar as the explanation of Supplementary Material IV in Ref. \cite{PhysRevResearch.5.023010}. Further, there are subharmonic and incommensurate responses of the revivals to 
driving. 

The main contribution of our work is the discovery of the conditions under which 
the Floquet scarring states emerge. The general conditions were first obtained
through a systematic probe of the parameter space defining the driving signal, 
revealing that these states are the result of a synchrony between the static 
detuning and the driving frequency. An application of the degenerate Floquet 
perturbation theory allowed us to analytically derive the emergence conditions. 
The theoretical analysis revealed that the Floquet scarring states originate 
from the resonances between these degenerate Fock base states that can 
be connected through a {\it one hopping process}. The resonances 
are induced by the first-order perturbation effect, weakening the constraint in 
the unperturbed dynamics. 

Floquet scarring states are of fundamental importance to many-body physics with 
significant applications in quantum control and engineering. Our work provides a 
stepping stone for further analyzing the breakdown of the ETH in solid-state systems 
and a more rigorous understanding of the Floquet scarring states. 

\section*{Acknowledgments}
This work was supported by the Air Force Office of Scientific Research under Grant
No.~FA9550-21-1-0186 and by the Office of Naval Research under Grant 
No.~N00014-24-1-2548.

\appendix

\section{Quantum dynamical evolution and related physical quantities} \label{appendix:A}

\subsection{Quantum evolution dynamics}

We reduce the dimension of the Hamiltonian Hilbert space following the method in 
Ref.~\cite{scherg2021observing}. For fixed numbers of spin-up ($N_{\uparrow}$) and 
spin-down ($N_{\downarrow}$) fermions in a lattice of $L$ sites, the number of spin 
$\sigma$ bases is 
\begin{equation}
    d_{\sigma} = \begin{pmatrix}
    L \\ N_{\sigma}
    \end{pmatrix}.
\end{equation} 
Denoting the occupation sites of the spin-up and spin-down fermions as 
$\left\{i_1,i_2,\cdots i_{N_{\uparrow}}, \right\}$ and 
$\left\{j_1,j_2,\cdots, j_{N_{\downarrow}} \right\}$, respectively, we obtain the 
typical number state as 
\begin{equation}
    \ket{\psi} = \hat{c}_{i_1,\uparrow}\hat{c}_{i_2,\uparrow}\cdots \hat{c}_{i_{N_{\uparrow}},\uparrow}\hat{c}_{j_1,\downarrow}\hat{c}_{j_2,\downarrow}\cdots \hat{c}_{j_{N_{\downarrow}},\downarrow} \ket{0}.
\end{equation}
The state can be represented by a pair of tuples 
$(\alpha,\beta)\equiv ((i_1,i_2,\cdots i_{N_{\uparrow}}),(j_1,j_2,\cdots,j_{N_{\downarrow}}))$
with the ordering $1 \le i_1< i_2< \cdots < i_{N_{\uparrow}} \le L$ and 
$1 \le j_1< j_2< \cdots < j_{N_{\downarrow}} \le L$. The number of full basis is 
thus $d_{\uparrow}\times d_{\downarrow}$ and a state is given by
\begin{equation}
    \ket{\psi} = \sum_{\alpha,\beta} \ket{\alpha,\beta} \braket{\alpha,\beta|\psi} \equiv \sum_{\alpha,\beta} M^{(\psi)}_{\alpha \beta}\ket{\alpha,\beta},
\end{equation}
where $M^{(\psi)}$ is a $d_{\uparrow}\times d_{\downarrow}$ matrix, and 
$\ket{\alpha,\beta}$ is the full basis corresponding to the tuple pair 
$(\alpha,\beta)$. The Hamiltonian becomes
\begin{equation}
    H = H_{\uparrow}^{\rm hop} \otimes \mathds{1}_{\downarrow} + \mathds{1}_{\uparrow} \otimes H_{\downarrow}^{\rm hop} + H^{\rm diag},
\end{equation}
where $\mathds{1}_{\sigma}$ is the $d_{\sigma}\times d_{\sigma}$ unit matrix, 
\begin{align} \nonumber
H_{\sigma}^{\rm hop} = \sum_i \hat{c}^{\dag}_{i,\sigma}\hat{c}_{i+1,\sigma}+{\rm h.c.} 
\end{align}
is the $d_{\sigma}\times d_{\sigma}$ matrix, and $H^{\rm diag}$ is a 
$d_{\uparrow}d_{\downarrow} \times d_{\uparrow}d_{\downarrow}$ diagonal matrix. 
Defining the $d_{\uparrow} \times d_{\downarrow}$ matrix 
$F \equiv {\rm diag}(H^{\rm diag})$ with the elements
\begin{equation}
    F_{\alpha \beta} = \left(\sum_{k=1}^{N_{\uparrow}} i_k + \sum_{k=1}^{N_{\downarrow}} j_k\right)\Delta +U N_d,
\end{equation}
where 
\begin{align} \nonumber
N_d = |(i_1,i_2,\cdots i_{N_{\uparrow}})\cap (j_1,j_2,\cdots, j_{N_{\downarrow}})| 
\end{align}
is the number of the doublons, we obtain the Schr\"{o}dinger equation as
\begin{align}
    \mathrm  i \sum_{\alpha,\beta} \frac{\partial M^{(\psi)}_{\alpha \beta}}{\partial t}\ket{\alpha,\beta} = & (H_{\uparrow}^{\rm hop} \otimes \mathds{1}_{\downarrow} + \mathds{1}_{\uparrow} \otimes H_{\downarrow}^{\rm hop} + F) \notag \\
    & \cdot \sum_{\alpha,\beta} M^{(\psi)}_{\alpha \beta} \ket{\alpha,\beta},
\end{align}
i.e.,
\begin{equation}
    \mathrm i \partial M^{(\psi)}/\partial t = H_{\uparrow}^{\rm hop}M^{(\psi)}+M^{(\psi)} H_{\downarrow}^{\rm hop}+F\circ M^{(\psi)},
\end{equation}
where $\circ$ represents the element-by-element multiplication (Hadamard product). An 
application of the Trotter-Suzuki decomposition stipulates that the dynamical 
evolution of the initial state is described by
\begin{align}
     M^{(\psi)}(t+\delta t) \approx e^{-\mathrm i \delta t \circ F} \circ e^{-\mathrm i \delta t H_{\uparrow}^{\rm hop}} M^{(\psi)}(t) e^{-\mathrm i \delta t H_{\downarrow}^{\rm hop}},
\end{align}
where the matrices $F$, $H_{\uparrow}^{\rm hop}$, and $H_{\downarrow}^{\rm hop}$ are 
all time-dependent and $e^{-\mathrm i \delta t \circ F}$ is the element-wise 
exponentiation. As a result, the matrix computation has been reduced from 
$d_{\uparrow}d_{\downarrow} \times d_{\uparrow}d_{\downarrow}$ dimension to 
$d_{\uparrow}\times d_{\downarrow}$ dimension.

\subsection{Bipartite von Neumann entanglement entropy and error analysis}
\begin{figure}[ht!]
\centering
\includegraphics[width=\linewidth]{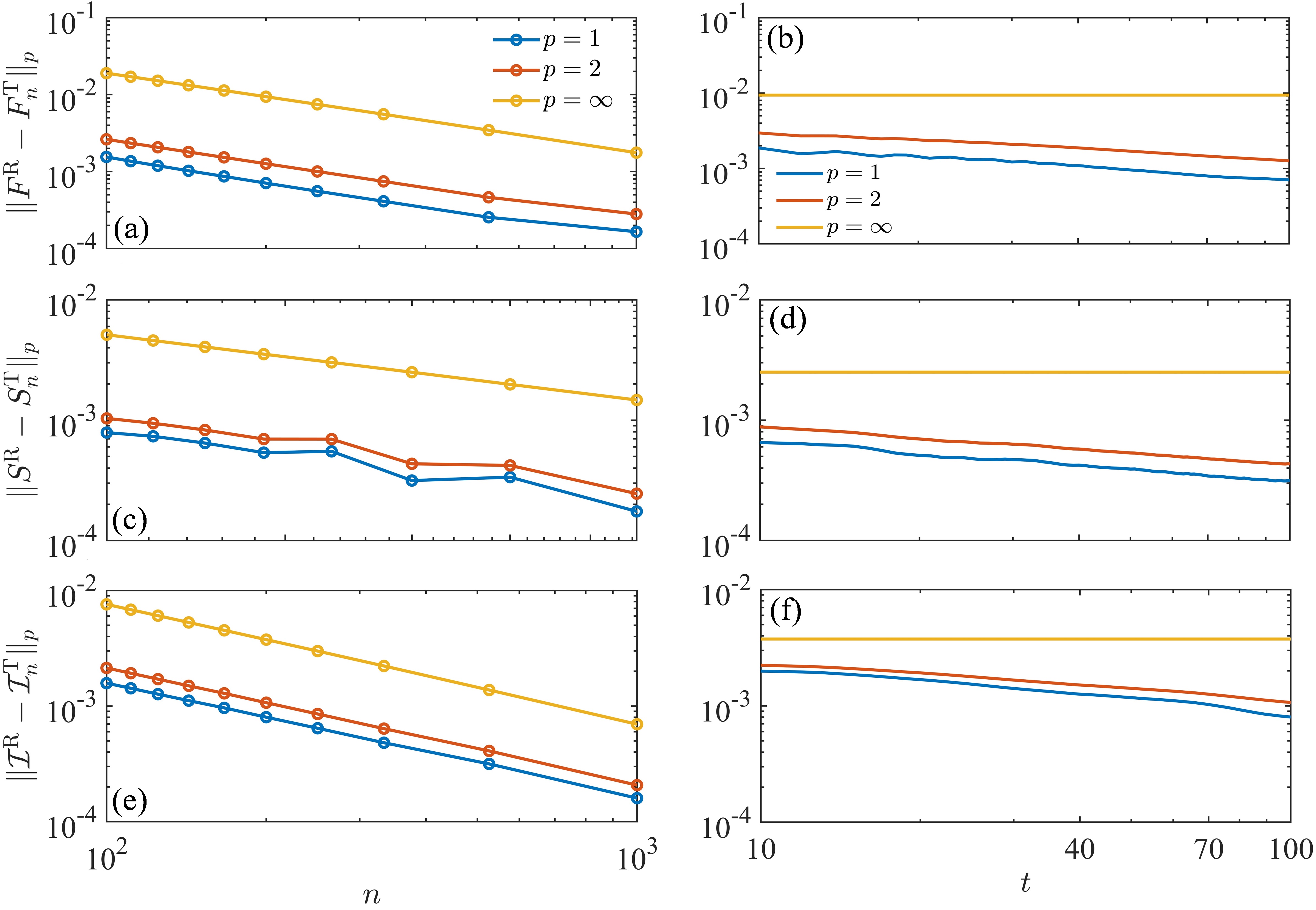}
\caption{Error estimates for Trotter-Suzuki decomposition. The exact values are 
calculated by the fourth order Runge-Kutta method. Shown are the standard 
$\mathcal{L}^p$-norm of (a-b) fidelity $F$, (c-d) bipartite entanglement entropy 
$S_{L/2}$, and (e-f) imbalance $\mathcal{I}$ as the function of (a,c,e) the Trotter 
steps $n$ or (b,d,f) time $t$.}
\label{fig:appendix_1}
\end{figure}

The basis numbers for the left and right half-chain are $d_l$ and $d_r$, respectively.
A typical quantum state is 
\begin{equation}
    \ket{\psi} = \sum_{l,r} \psi_{lr} \ket{l} \otimes \ket{r},
\end{equation}
where $\psi_{lr}$ is the element of the $d_l\times d_r$ matrix $\psi$, $\ket{l}$ and 
$\ket{r}$ are the bases of the left and right half-chain, respectively. The reduced 
density matrix is 
\begin{align}
    \rho_l & = {\rm tr}_r \ket{\psi}\bra{\psi} \notag \\
    & = \sum_{r'} \braket{r'|\psi}\braket{\psi|r'} \notag \\
    & = \psi\psi^{\dag},
\end{align}
and similarly $\rho_r = (\psi^{\dag}\psi)^{T}$. Using the singular value 
decomposition, we obtain the matrix $\psi$ as 
\begin{equation}
    \psi = A\Sigma B^{\dag},
\end{equation}
where $\Sigma$ is a $d_l\times d_r$ diagonal matrix, $A$ and $B$ are $d_l\times d_l$ 
and $d_r\times d_r$ unitary matrices, respectively. When the lattice number $L$ is 
even, we have $d_l = d_r = 2^L$ and the bipartite von Neumann entanglement entropy is 
\begin{equation}
    S_{L/2} = S_l = S_r =-\sum_{i=1}^{d_l} \Sigma_i^2\ln\Sigma_i^2.
\end{equation}
The Trotter-Suzuki decomposition leads to error accumulation, but the error decreases 
with increased time-steps $n$ in per time unit $\tau$. The error can be quantified by
The standard $\mathcal{L}^p$-norm
\begin{equation}
    \Vert \mathcal{O}^{\rm R} - \mathcal{O}_n^{\rm T} \Vert_p = \left(\int_0^{t} |\mathcal{O}^{\rm R}(t) - \mathcal{O}_n^{\rm T}(t)|^p dt \right)^{1/p}
\end{equation}
with $p=1,2,...,\infty$, where $\mathcal{O}$ is some physical quantity, 
$\mathcal{O}^{\rm R}$ represents the exact value calculated by the fourth order 
Runge-Kutta method, and $\mathcal{O}^{\rm T}_n$ is the value calculated by the 
$n$-steps Trotter-Suzuki decomposition. Specifically, $p=1$ means the average 
difference between $\mathcal{O}^{\rm T}_n$ and $\mathcal{O}^{\rm R}$ and $=\infty$ 
with 
\begin{align} \nonumber
\Vert \mathcal{O}^{\rm R} - \mathcal{O}_n^{\rm T} \Vert_{\infty} = \max{(|\mathcal{O}^{\rm R}(t) - \mathcal{O}_n^{\rm T}(t)|)} 
\end{align}
means the largest difference between them. Figures~\ref{fig:appendix_1}(a-b) show 
$\mathcal{L}^p$-norms with $p=1,2,\infty$ of the fidelity $F$ for different 
time step $n$ with the fixed integration upper bound $t = 100\tau$, and for different 
upper bound $t$ for a fixed time-steps $n=200$, respectively. 
Figures~\ref{fig:appendix_1}(c-d) and \ref{fig:appendix_1}(e-f), respectively, 
display the corresponding $\mathcal{L}^p$-norms for the bipartite von Neumann 
entanglement entropy $S_{L/2}$ and the imbalance $\mathcal{I} = (N_o-N_e)/(N_o+N_e)$ 
on the even and odd sublattices. In an approximate sense, the $\mathcal{L}^p$-norm 
approaches zero as $1/n$, and decreases slightly for increasing time.

\begin{figure} [t!]
\includegraphics[width=1\linewidth]{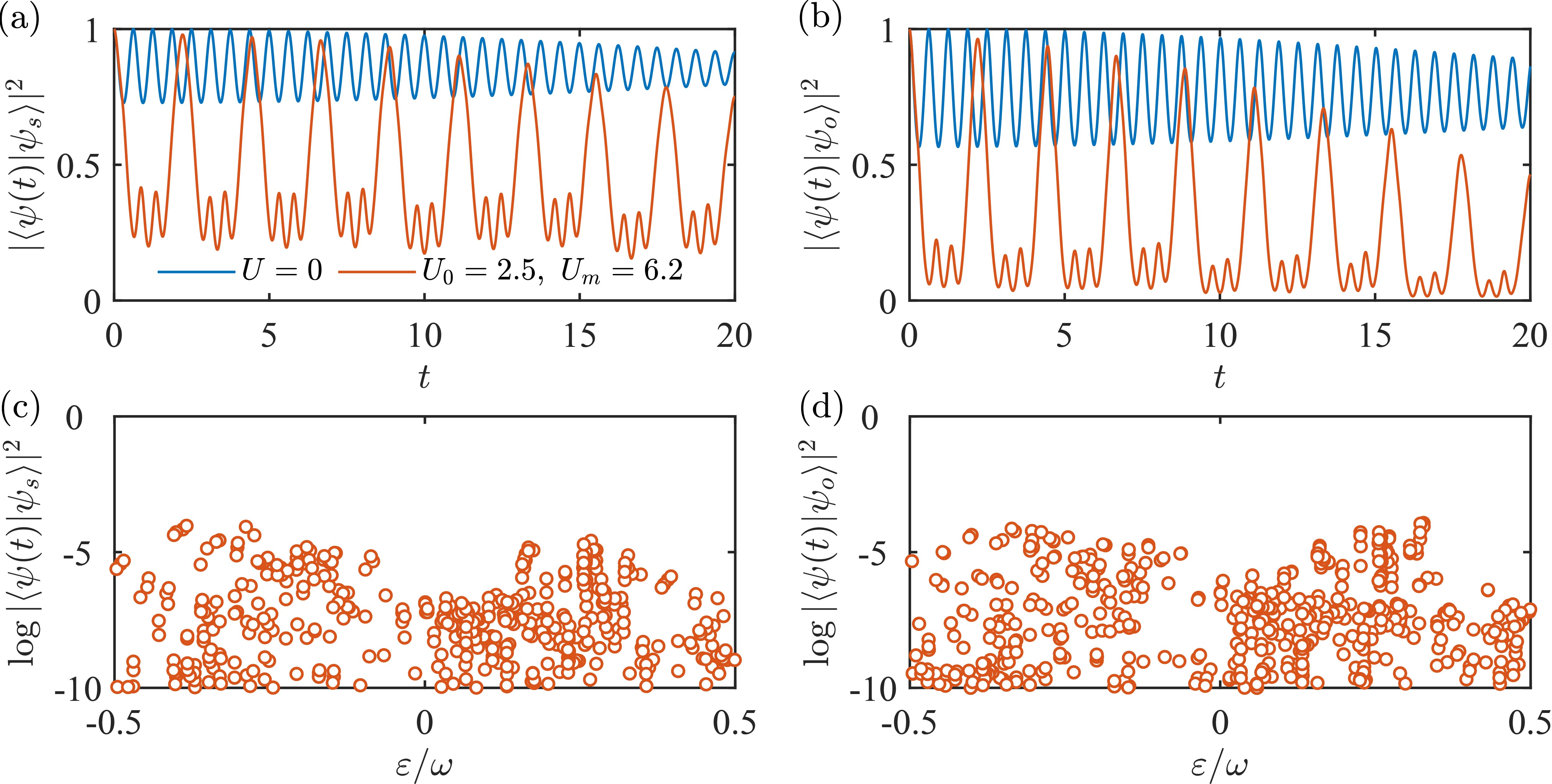}
\caption{Properties of the transition state encircled by black curves in 
Fig.~\ref{fig:fid_ave}(a). The parameters are $(U_0,U_m,\omega)=(2.5,6.2,2\sqrt{2})$. 
(a-b) Dynamics of the wave function fidelity in a quench process from the initial 
state (a) $\ket{\psi_s}$ or (b) $\ket{\psi_{th}}$. Wannier-Stark localization with 
$U=0$ is colored in blue and the transition state is colored in orange. 
(c-d) The overlap of the Floquet eigenstates with (c) $\ket{\psi_s}$ or 
(d) $\ket{\psi_{th}}$.}
\label{fig:appendix_2}
\end{figure}

\section{Wannier-Stark localization} \label{appendix:B}
For a noninteracting system with $U = 0$, the Hamiltonian can be diagonalized as~\cite{van2019bloch}
\begin{equation}
    H = \sum_{m,\sigma = \uparrow,\downarrow} \Delta m \hat{b}_{m,\sigma}^{\dag}\hat{b}_{m,\sigma}+{\rm h.c.},
\end{equation}
by the transformation:
\begin{equation}
\hat{b}_m = \sum_{j,\sigma = \uparrow,\downarrow} \mathcal{J}_{j-m}(2J/\Delta)\hat{c}_{j,\sigma},
\end{equation}
where $\mathcal{J}_n$ is the Bessel function of the first kind. Since 
$|\mathcal{J}_n (2J/\Delta)|<e^{-|n|}$ for $2J/\Delta \ll n$, all the eigenstates 
are localized for any $\Delta \neq 0$ - the phenomenon of called Wannier-Stark 
localization~\cite{wannier1960wave}. More specifically, each eigenstate 
is localized about site $m$ with an inverse localization length
\begin{align} \nonumber
\xi^{-1} \approx 2\sinh^{-1}(\Delta/2J) 
\end{align}
and exhibits Bloch oscillations~\cite{dahan1996bloch} with the characteristic 
period $T = h/\Delta = 2\pi\tau/\Delta$ in our units. The wave function fidelity 
oscillates about a high value, as shown in blue in Figs.~\ref{fig:appendix_2}(a) and 
\ref{fig:appendix_2}(b). This is a manifestation of Bloch oscillations of the period 
$T \approx 0.628$, in consistence with the theoretical result. 

In Figs.~\ref{fig:appendix_2}(a-d), the orange represents the case in the regions 
encircled by the black curves in Fig.~\ref{fig:fid_ave}(a): $\omega = 2\sqrt{2}$, 
$U_0 = 2.5$, and $U_m=6.2$. The fidelity oscillates about a value that decays slowly 
over time. It does not decrease to zero and so does not indicate a revival behavior. 
In addition, there is no intrinsic difference between the initial states 
$\ket{\psi_s}$ and $\ket{\psi_{th}}$, for both the quantum fidelity 
[Figs.~\ref{fig:appendix_2}(a) and \ref{fig:appendix_2}(b)] and the overlap of 
Floquet eigenstates with the initial states [Figs.~\ref{fig:appendix_2}(c) and 
\ref{fig:appendix_2}(d)]. Especially in Fig.~\ref{fig:appendix_2}(c), the tower 
structure and the anomalously high overlap with $\ket{\psi_s}$ do not exist. While 
both the average fidelity $\langle F_s \rangle_t$ from $\ket{\psi_s}$ and the 
relative discrepancy $\varrho$ are high, none of the above characteristics are 
consistent with the scarring dynamics. In this regard, these regions encircled by 
black curves in Figs.~\ref{fig:fid_ave}(a) and \ref{fig:fid_ave}(b) correspond to 
the transition states from Wannier-Stark localization to the Floquet scarring phase.

\section{The Floquet perturbation theory} \label{appendix:C}
The Hamiltonian $H(t)=H_0(t)+ V$ has period $T$, where $V$ is the time-independent perturbation term. Assuming that $H_0(t)$ commutes with itself
at different times, its eigenstates $\ket{m}$ are time-independent in the specific basis, as the result of $H_0(t)\ket{m} = E_m(t)\ket{m}$ and $\braket{q|m} = \delta_{qm}$. And we also assume that $V$ is completely off-diagonal in this basis, i.e., $\braket{m|V|m}=0$ for all $\ket{m}$. 

The Floquet modes $\ket{m(t)}$ of $H(t)$ satisfy the Schr{\"o}dinger equation:
\begin{equation} \label{eq:C1}
    \mathrm i \frac{\partial \ket{m(t)}}{\partial t} = H(t)\ket{m(t)},
\end{equation}
and
\begin{equation} \label{eq:C2}
    \ket{m(T)} = e^{-\mathrm i \varepsilon_m}\ket{m(0)},
\end{equation}
where $\varepsilon_m$ are quasienergies of $H(t)$, and $\varepsilon_m$ are eigenvalues of Floquet Hamiltonian $H_{\rm F}$: $H_{\rm F}\ket{m}=\varepsilon_m\ket{m}$. When $t=0$, the Floquet modes $\ket{m(0)}$ are referred to as Floquet eigenstates, which are indeed equivalent to the eigenstates $\ket{m}$.  For $V=0$, we have $\ket{m(t)}=e^{-\mathrm i \int_0^t dt' E_m(t')}\ket{m}$, and $e^{-\mathrm i \varepsilon_m}=e^{-\mathrm i \int_0^T dt E_m(t)}$.

\begin{figure} [t!]
\includegraphics[width=\linewidth]{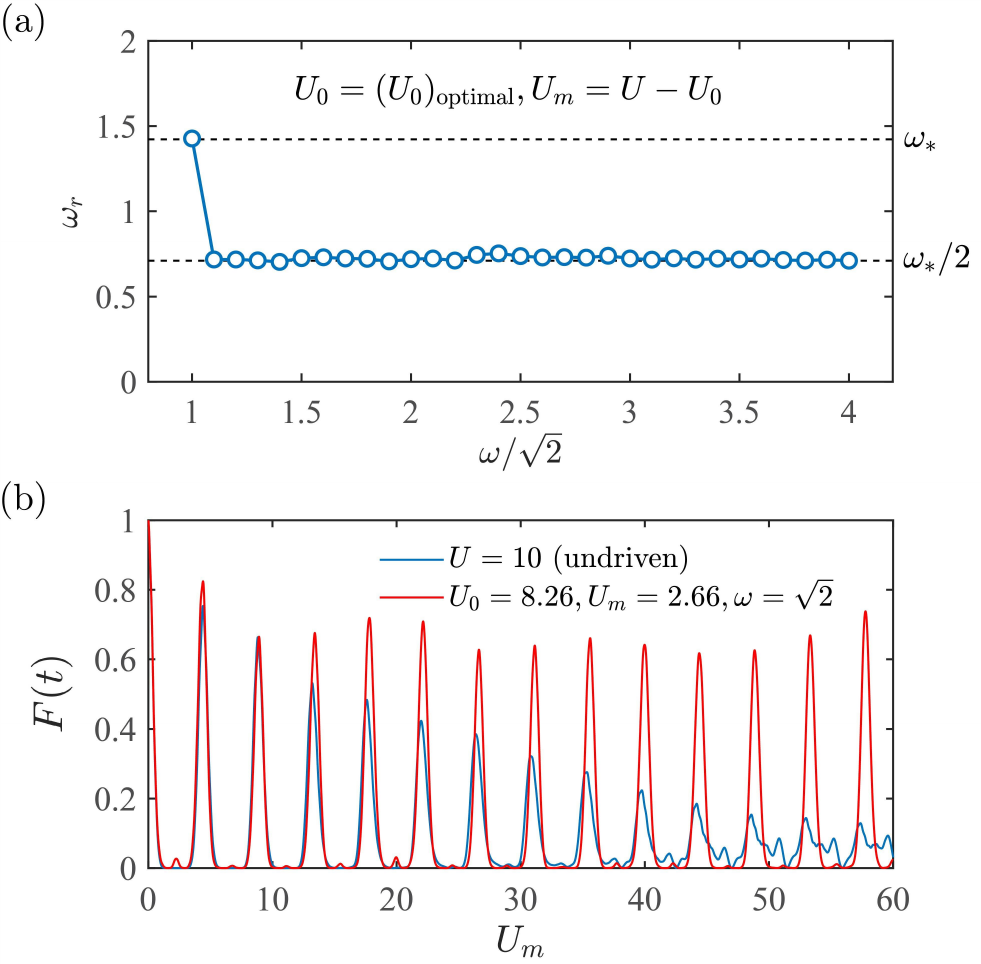}
\caption{Period doubling of quantum revival under square-wave drive. (a) Period 
doubling occurred for $U_0 =(U_0)_{\rm optimal}$ and $U_m = U-U_0$, for $U = 10$. 
(b) Driven quantum revival from $\ket{\psi_s}$, which is enhanced and stabilized 
by the square-wave driving, for optimal parameter set 
$(U_0,U_m,\omega) = (8.26,2.66,\sqrt{2})$.}
\label{fig:PD}
\end{figure}

For small $V$, the Floquet modes $\ket{m(t)}$ can be expanded in terms of the unperturbed eigenstates:
\begin{equation} \label{eq:C3}
    \ket{m(t)} = \sum_{q} c_q(t) e^{-\mathrm i \int_0^t dt' E_q(t')} \ket{q},
\end{equation}
where $c_m(t) \simeq 1$ for all $t$, and $c_q(t)$ is of the order $V$ for all $q\neq m$ and all $t$. Plugging Eq. (\ref{eq:C3}) into the Schr{\"o}dinger equation, we can simply to obtain
\begin{equation*}
    \mathrm i \sum_{q} \frac{d c_q(t)}{dt} e^{-\mathrm i \int_0^t dt' E_q(t')} \ket{q} = V \sum_{q} c_q(t) e^{-\mathrm i \int_0^t dt' E_q(t')} \ket{q},
\end{equation*}
then take the inner product with $\bra{m}$, we have
\begin{align} \label{eq:C4}
       \mathrm i \frac{d c_m(t)}{dt} &= c_m(t) \braket{m|V|m} \notag \\ 
       & \quad + \sum_{q \neq m} c_q(t) e^{\mathrm i \int_0^t dt' [E_m(t') - E_q(t')]} \braket{m|V|q}.
\end{align}
Since $\braket{m|V|q}$ and $c_q(t)$ are of the order $V$, their product in the sum represents the second-order term in $V$ that can be disregard. And $\braket{m|V|m} = 0$, we have $d c_m(t)/dt=0$. Thus $c_m(t)$ can be chosen as 1 for all $t$, and then
\begin{align} \label{eq:slight}
    \ket{m(t)} =  e^{-\mathrm i \int_0^t dt' E_m(t')} \ket{m} + \sum_{q\neq m} c_q(t) e^{-\mathrm i \int_0^t dt' E_q(t')} \ket{q},
\end{align}
where $c_q(t)$ is of the order $V$ for all $q\neq m$ and all $t$.

Taking the inner product with $\bra{q(t)}$ and integrating the Schr\"{o}dinger 
equation (\ref{eq:C1}) from $t=0$ to $t=T$, we obtain
\begin{align} \label{eq:cqT_1}
c_q(T)=c_q(0)-\mathrm i\braket{q|V|m}\int_0^T dt e^{\mathrm i\int_0^t dt' [E_q(t')-E_m(t')]}.
\end{align}
In addition, utilizing the relation (\ref{eq:C2})
for all $q\neq m$, we get 
\begin{align} \label{eq:cqT_2}
c_q(T) = e^{\mathrm  i \int_0^T dt [E_q(t)-E_m(t)]} c_q(0).
\end{align} 
Combining Eqs.~(\ref{eq:cqT_1}) and (\ref{eq:cqT_2}), we have
\begin{equation} \label{eq:undegenrated}
    c_q(0) = -\mathrm i \braket{q|V|m} \frac{\int_0^T dt e^{\mathrm i\int_0^t dt' [E_q(t')-E_m(t')]}}{e^{\mathrm i \int_0^T dt [E_q(t)-E_m(t)]}-1}.
\end{equation}
The analysis so far holds for nondegenerate states. It breaks down when degeneracy 
occurs under the condition:
\begin{equation} \label{eq:degenerated}
    e^{\mathrm i \int_0^T dt [E_q(t)-E_m(t)]} = 1.
\end{equation}

Suppose that there are $p$ states satisfying the 
condition (\ref{eq:degenerated}) with $\ket{m}$, denoted as $\ket{m_i}$ with $i=1,2,\cdots,p$ and $\ket{m} \equiv \ket{m_0}$. Ignoring all the
other states of the system for the moment, the Floquet mode $\ket{m_i(t)}$ now is
\begin{equation} \label{eq:psi}
    \ket{m_i(t)} = \sum_{j = 0}^p c_j(t) e^{-\mathrm i \int_0^t dt' E_j(t')} \ket{m_j}
\end{equation}
for $i = 0,1,\cdots,p$, where all the $c_j(t)$'s are of order one (instead of order $V$). Now Eq. (\ref{eq:C4}) is
\begin{equation} \label{eq:dt}
    \mathrm i \frac{d c_i(t)}{dt} = \sum_{j \neq i} c_j(t) e^{\mathrm i \int_0^t dt' [E_i(t') - E_j(t')]} \braket{m_i|V|m_j},
\end{equation}
the sum term is no longer a second-order term in $V$. To first order of $V$, we can replace $c_j(t)$ by $c_j(0)$ on the right-hand side of Eq. (\ref{eq:dt}), then upon integrating from $t=0$ to $t=T$, we have
\begin{align*}
    c_i(T) & =  c_i(0) -\mathrm i \sum_{j \neq i} \braket{m_i|V|m_j} c_j(0) \\
    & \quad \times \int_0^T dt e^{\mathrm i \int_0^t dt' [E_i(t') - E_j(t')]}.
\end{align*}
This can be written as matrix form
\begin{equation}
    c(T) = (I-\mathrm i M)\cdot c(0),
\end{equation}
where $c(t) = [c_0(t),c_1(t),\cdots,c_p(t)]^T$ and the $(p+1)\times (p+1)$ matrix $M$ has 
the elements
\begin{equation}
M_{ij} = \braket{m_i|V|m_j}\int_0^T dt e^{\mathrm i\int_0^t dt' [E_{i}(t')-E_{j}(t')]}.
\end{equation}
Let the eigenvalues of $M$ be $\varsigma_i$ with $i=0,1,\cdots,p$. The corresponding
eigenstates are $c(T) = e^{-\mathrm i \varsigma_i} c(0)$. The Floquet modes $\ket{m_i(t)}$ satisfy the condition 
\begin{align} \nonumber
\ket{m_i(T)} = e^{-\mathrm i \varepsilon_i T}\ket{m_i(0)}.
\end{align}
Thus the Floquet quasienergies are then given by
\begin{equation}
    e^{-\mathrm i \varepsilon_i T} = e^{-\mathrm i \varsigma_i-\mathrm i \int_0^T dt E_i(t)},
\end{equation}
and the Floquet Hamiltonian is
\begin{equation}
    (H_{\rm F})_{ij} = \frac{M_{ij}}{T}.
\end{equation}

\section{Robust period-doubling} \label{appendix:D}
In contrast to the tunable responses, there is a robust period-doubling phenomenon 
relating the driven and undriven revival periods: $T_r = 2T_*$ for 
$U_0 = (U_0)_{\rm optimal}$ and $U_m = U-U_0$, as exemplified in Fig.~\ref{fig:2}(a). 
Figure~\ref{fig:PD}(a) shows such a phenomenon, for $\omega_r \approx \omega_*/2$ 
over a wide range of $\omega$. For $\omega = \sqrt{2}$, there is a harmonic response: 
$T_r \approx T_*$. In this case, we have identified an overall optimal parameter set 
$(\omega,U_0,U_m)=(\sqrt{2},8.26,2.66)$ for 31 values of the driving frequency, in 
which the quantum revival is greatly enhanced and stabilized by periodic driving, 
especially for a long-time evolution, as shown in Fig.~\ref{fig:PD}(b). The optimal 
driving frequency is close to the undriven revival frequency: 
$\omega_{\rm optimal} \approx \omega_*$, and the driven revival frequency is close 
to the optimal driving frequency: $\omega_r \approx \omega_{\rm optimal}$ 
(the harmonic response).

\bibliography{QMBS}
\end{document}